\documentclass{aa} \usepackage{graphics} 
\begin{document} 
\thesaurus{13.09.1, 13.21.1, 11.12.2, 11.19.7} 
\title
{Far infrared and Ultraviolet emissions of individual galaxies 
at z=0: selection effects on the estimate of the dust extinction}

\author {V. Buat\inst{1,2}, J. Donas\inst{1}, B.Milliard\inst{1}, C. 
Xu\inst{3}}
\institute { IGRAP, Laboratoire
d'Astronomie Spatiale du CNRS, BP 8, 13376 Marseille Cedex 12, France \and
Laboratoire des interactions photons-mati\`ere, Facult\'e des Sciences de 
Saint
J\'er\^ome, 13397 Marseille Cedex 13, France\and California Institute of 
Technology, Pasadena, CA 91125, USA}
\date{ Received; accepted ....  }
\offprints{V. Buat}
\mail{buat@astrsp-mrs.fr}
\titlerunning{FIR and UV properties of galaxies at z=0 }
\authorrunning{Buat et al.}
\maketitle 
\begin{abstract} 
We have cross-correlated Far Infrared (IRAS) and UV (FOCA) observations of 
galaxies to construct a sample of FIR selected galaxies with a UV 
observation at 0.2 $\mu$m.\\
 The FIR and UV properties of this sample are compared to the mean properties
of the local Universe deduced from the luminosity distributions at both
wavelengths.  Almost all the galaxies of our sample have a FIR to UV flux ratio
larger than the ratio of the FIR and UV luminosity densities, this effect
becoming worse as the galaxies become brighter:  the increase of the UV (0.2
$\mu$m) extinction is about 0.5 mag per decade of FIR (60 $\mu$m) luminosity.\\
Quantitative star formation rates are estimated by adding the contribution of 
the FIR and UV emissions. They are found consistent with the corrections for 
extinction deduced from the FIR to UV flux ratio.
A total local volume-average star formation rate is calculated by summing the 
contribution of the FIR and UV wavelengths bands. Each band contributes for an 
almost similar amount to the total star formation rate with 
$\rm \rho_{SFR} = 0.03 \pm 0.01 ~h \cdot
M\odot/yr/Mpc^3$ at z=0. This is equivalent to a global extinction of 0.75 mag 
to 
apply to the local luminosity density at 0.2$\mu$m.\\
The trend of a larger FIR to UV flux ratio for a larger FIR luminosity found
for our sample of nearby galaxies is extended and amplified toward the very 
large FIR luminosities
when we consider the galaxies detected by ISOCAM in a CFRS field and the Ultra
Luminous Infrared Galaxies at low and high redshift.  A UV extinction is
tentatively estimated for these objects.

\keywords {Infrared: galaxies--Ultraviolet: galaxies--Galaxies: luminosity 
function--Galaxies: statistics} 
\end{abstract}

\section{Introduction}

The problem of internal dust extinction in galaxies is difficult whereas its
estimate is crucial for understanding the evolution of the Star Formation Rate
from high redshift to now.  At high z the emission observed in the visible
corresponds to the UV rest frame where the effects of the dust extinction can
be dramatic.  For example the shape of the Madau plot (e.g.  Madau et al.
\cite{madau}) depends a lot on the extinction adopted as a function of the
redshift.  Since the work of Calzetti, Kinney and collaborators from IUE data
(e.g.  Kinney et al.  \cite{kinney}, Calzetti et al.  \cite{calzkinn}) the
slope of the UV continuum in the range 1200-2600 \AA ($\rm f_{\lambda}\propto
\lambda^{\beta}$) has been identified as a powerful indicator of the dust
extinction.  The reason is that only short lived stars contribute substantially 
to the
emission in this wavelength range and the intrinsic shape of the spectrum is
only sensitive to the recent star formation history.  For example $\beta$
reaches a steady  value $\sim -2$ as soon as the star formation rate has
been constant for some $10^7$ years (Calzetti et al.  \cite{calzkinn}). An
instantaneous starburst represents the most extreme case of a steep spectrum 
with $\beta$ equal to
-2.7 ( Meurer et al.  \cite{meurheckII}).  Once an intrinsic value for $\beta$ 
is 
adopted any deviation from this 
value is interpreted in terms of dust extinction which flattens the intrinsic
slope.  At high redshift $\beta$ is observable in the visible wavelength range
and this has conferred a large interest to this approach.

Nevertheless, at least two difficulties arise when using this method:  on the
one hand the choice of an intrinsic UV slope $\beta$ (i.e.  of a star formation
history) can modify substantially the amount of extinction and has led to some
discrepancies in the estimate of the extinction which can reach 1-2 mag in UV
(Pettini et al.  \cite{pettini}, Meurer et al.  \cite{meurheckI}, Steidel et
al.  \cite{steidel}, Calzetti \cite{calzetti}); on the other hand if the
deviation of $\beta$ from its intrinsic value is indubitably a dust extinction
tracer, deriving a quantitative value of the extinction from this deviation is
difficult due to the various unknown factors like geometry or dust properties
intervening in the estimate of the extinction (e.g.  Calzetti et al.
\cite{calzkinn}).  The quantification of the extinction is easier on nearby
galaxies which can be  obviously used as templates.  The empirical approach of
Calzetti and collaborators ( Calzetti et al.  \cite{calzkinn}, Kinney et al.
{\cite{kinncalz}, Calzetti \cite{calzetti}) had the advantage of providing  a
global attenuation curve for starburst galaxies accounting for geometrical
effects in a statistical way.

 Another powerful approach lies in making  global energetic considerations.  
Indeed,
for nearby templates, it is possible to perform a total energetic budget since
the dust emission of these galaxies is almost always known from IRAS 
observations.
Such considerations have led to quantitative estimates of the extinction (Buat 
\&
Xu \cite{buxu}, Meurer et al.  \cite{meurheck}).  Recently, Meurer et al.
(\cite{meurheck}) have related the FIR to UV flux ratio and the UV slope 
$\beta$
to the extinction at 1600 \AA.  Their unreddened UV spectrum has a slope of
-2.23 intermediate between a constant star formation rate and an instantaneous
burst.

Uncertainties about the UV extinction are already present at low redshifts.
Meurer et al.  (\cite{meurheck}) find an extinction around 1.8 mag at 1600 $\rm
\AA$ for the starburst templates observed by IUE whereas we find 1.3 mag at 2000
$\rm \AA$ for a sample of nearby starburst galaxies (Buat \& Burgarella
\cite{bubu}) and around 0.8 mag for more quiescent disk galaxies (Buat \& Xu
\cite{buxu}).  The difference is likely to be due at least in part to the
properties of the individual galaxies used for these studies but also to
different assumptions about the dust absorption as it will be discussed below.

 We need to know how to correct individual galaxies for extinction but also how
the properties of these individual cases can be extrapolated to the entire
population of galaxies.  This problem is especially important at high z since as
we go farther only the brightest objects become visible.  With the availability
of the luminosity functions at various wavelengths we have now the possibility
to test if the results deduced from the properties of individual galaxies are
representative of the mean characteristics deduced from the local luminosity
functions.  \\

 The basic idea of this paper is to compare the FIR (60 and 100
$\mu$m) and UV(0.2$\mu$m) of a sample of nearby galaxies for which selection
biases are well known.  The sample will be FIR selected and the aim is to
study how much these individual galaxies thus selected trace the mean properties
of the local universe.  After a presentation of our IRAS/FOCA sample (section 
2), we discuss the FIR and UV properties of the individual galaxies in terms of 
extinction and selection biases in section 3. The section 4 is a comparison with 
the luminosity functions at both wavelengths. In section 5 we derive 
quantitative star formation rates from the FIR and UV emissions. Endly, in 
section 6 we compare 
 the FIR and UV properties of Ultra Luminous Infrared Galaxies
both at low and high redshift and  of the ISOCAM detections of
intermediate redshift galaxies in a CFRS field with those of our IRAS/FOCA 
sample of
nearby galaxies.\\

\section{The IRAS/FOCA sample}
\begin{table*}
\caption[]{Fields  at 0.2 $\mu$m observed by the FOCA experiment with the 
coordinates 
of the field center (guide star), the total exposure time in each field and the  
 size of the field (diameter)}
\begin{flushleft}
\begin{tabular}{lllrll} 

\hline 
Field number&$\alpha$ (1950)&$\delta$(1950) &Exposure &time (s)& Field diameter 
\\
& h m s&d m s &FOCA1000&/FOCA1500&deg\\ 
\hline 
12&00 34 43.9&+40 03 27&1200&/~~&2.3\\
54&03 46 58.4&+22 05 37&600&/~~&2.3\\
10&08 17 26.1&+20 54 25&1400&/~~&2.3\\
51&08 48 36.0&+43 54 51&600&/800&2.3\\
18&08 51 57.6&+78 20 18&600&/~~&2.3\\
82&09 58 57.0&+69 01 41&2400&/~~&2.3\\
67&11 42 45.5&+20 10 03&600&/1600&2.3\\
33&11 59 04.1&+65 13 04&2000&/~~&2.3\\
81&12 19 29.2&+47 27 34&600&/~~&2.3\\
71&12 18 16.7&+15 49 06&~~&/800&1.5\\
50&12 25 09.4&+08 53 13&450&/~~&2.3\\
34&12 26 02.3&+12 23 39&1800&/~~&2.3\\
28&12 57 08.1&+28 20 06&3000&/~~&2.3\\
30&13 03 47.1&+29 17 48&3000&/1200&2.3\\
31&13 09 32.4&+28 07 52&1050&/~~&2.3\\
96&13 29 50.2&+47 29 28&~~&/1200&1.5\\
90&13 39 52.9&+28 37 38&300&/3600&2.3\\
29&14 01 04.2&+54 54 21&~~&/1200&1.5\\
91&15 16 01.9&+02 15 51&~~&/1200&1.5\\
89&15 36 52.8&+34 50 13&~~&/3600&1.5\\
36&16 39 16.7&+36 17 46&~~&/800&1.5\\
39&17 15 35.0&+43 11 21&~~&/1200&1.5\\
\hline  
\end{tabular} 
\end{flushleft} 
\end{table*}

\subsection{Construction of the sample of galaxies}

The FOCA balloon borne wide-field UV camera (Milliard et al. \cite{milliard}) 
has observed a cumulated sky surface of $\sim 100$ 
 square degrees in a 150$\rm  \AA$ wide band-pass
centered near 0.2 $\rm \mu m$. The camera (a 40-cm Cassegrain telescope with an 
image
intensifier coupled to a IIaO emulsion film) was operated in two modes, the
FOCA 1000 (f/2.6) and FOCA 1500 (f/3.8), which provide 2.3 $\deg$ field of 
view,
20 $\arcsec$ resolution, and 1.5 $\deg$ field of view, 12 $\arcsec$ resolution,
respectively. The typical limiting depth in one hour observing time is
 $\rm m (0.2 \mu m) =18.5$ where the  magnitude  is  defined by  
$\rm m (0.2 \mu m ) = -2.5 ~\log f-21.175 $ 
where the flux f is in $\rm erg/cm^2/s/\AA$ (Donas et
al. \cite{donas}). 
Here we have considered the 22 calibrated fields ($\sim$ 70 square degrees), in 
order to cross correlate
them with the observations of the IRAS satellite. Table 1 gives the coordinates
of the guide star (near the field center), the total exposure time and the size 
for  each field. 

The infrared objects of the IRAS Faint Sources Catalog (FSC) have been
associated to sources from other astronomical catalogs.  Such 
cross-correlations
are very useful to determine the nature of the sources detected.  Unfortunately
only a small proportion of UV sources have an identification in an other
catalog.  So we have chosen to start from the IR detections for which
much data are available and to search for their UV counterparts.  {\it 
Therefore
our sample will be FIR selected}.

For each UV field, we have extracted the FIR sources detected by IRAS at 60
and/or 100 microns and listed in the IRAS FSC.  This has been done using the
VIZIER facility of the Centre de Donn\'ees astronomiques de Strasbourg (CDS).
364 IR sources have been selected.\\ Since we are only interested by the
extragalactic targets we have kept only the objects associated to known galaxies
from the catalog of associations of the FSC.  102 from the 364 sources at 60 or
100 microns have been securely identified as galaxies.  We have only kept 
galaxies which are
not confused with a neighboured source present in cross-correlated catalogs\\ 
Then we have searched for a UV
source matching each FIR detection of a galaxy in a circle of 45 arcsec radius
centered on the IRAS coordinates.  94 galaxies have been identified both in UV
and FIR.   8 FIR sources identified as galaxies have no UV counterpart. Few 
cases of several UV sources present in the circular area have been judged 
doubtful and discarded.\\ 
To
avoid a contamination in the IR detection only galaxies with a cirrus flag lower
than or equal to 2 are selected as adviced in the IRAS Faint Source Catalog.  We
are then left with 80 galaxies with a UV measurement and with 8 galaxies
detected by IRAS and not identified in UV.\\ Endly we exclude nearby ellipticals
and S0 galaxies present in our sample since we are only concerned by star
forming galaxies.  Very extended galaxies like M101 or M51 are excluded from the
study since the photometry of these objects needs a special treatment.  Our
final sample contains 76 galaxies.

Complementary data necessary to the study of this sample like the optical
identification, the B magnitude, the distance modulus are taken from the LEDA
 and NED databases.  The fluxes are corrected for Galactic extinction using the 
Milky Way extinction curve of Pei (\cite{pei}). Throughout the paper h will be 
defined as $\rm H_0/100 ~km~ s^{-1} Mpc^{-1}$.
\begin{table*}
\caption[]{Galaxies detected in FIR with no or an uncertain UV detection. The B 
magnitude are taken from Yuan et al. \cite{yuan} and the CDS database.}
\begin{flushleft}
\begin{tabular}{lllllll} 

\hline 
IRAS name &$\rm f(60\mu m)$ &$\rm f(100\mu m)$& $\rm m_B$ &$\rm m(0.2\mu 
m)$&$\rm 
\log(F_{60}/F_{0.2})$ &$\rm a_{UV}$\\ 
        &Jky&Jky&mag &mag&&mag\\
\hline 
F11431+2037  &   0.329 &  $<0.72$&&$>18.5$&$>1.82$&$>3.8$\\
F12041+6519  &   0.206  & 0.60&17&17.2& 1.11:&2.3:\\
F12235+0914   &  0.260 & $< 0.94$&15&$>17.2$&$>1.21$&$>2.6$\\
F12242+0919   &  0.331 & $< 0.96$&14.75&$>17.2$&$>1.32$&$>2.7$\\
F12259+1141  &   0.266  &$< 1.92$&18.4&$>18.5$&$>1.72$&$>3.8$\\
F13041+2907  &   0.297  & 0.53&&$>18.5$&$>1.77$&$>3.6$\\

\hline  
\end{tabular} 
\end{flushleft} 
\end{table*}

\subsection{The  galaxies not detected at 0.2 $\mu$m}

8 galaxies detected by IRAS at least at 60 $\mu$m do not appear in our UV
frames.  For one of them (F15451+0132), the non detection is explained by the
fact that the galaxy is located on the edge of the UV image.  Another source,
F13038+2919, lies on the wings of the UV bright guide star (spectral type A3,
$\rm m (0.2\mu m)<7.3$).  The UV identification of F12041+6519 (identified as
MCG 11-15-022 in NED) at 60 arcsecs from the IRAS coordinates is quite 
uncertain
.  In table 2 are gathered the galaxies for which we can estimate an upper 
limit
for their UV flux and F12041+6519 whose identification is uncertain.   These
galaxies are faint even in FIR:  4 of them are only detected at 60 $\mu$m with 
a
very low flux.\\
{\it F13041+2907} and {\it F11431+2037} have no optical counterpart,  
{\it F13041+2907} is  also 
detected  at 1.4 GHz by FIRST (NED database).\\
{\it F12259+1141, F12235+0914} and {\it F12242+0919} are three galaxies located 
in  the 
Virgo cluster area and identified by Yuan et al. (\cite{yuan}):
{\it F12259+1141} (VCC1099) is a  faint  galaxy classified as 
dE; 
{\it F12242+0919} (VCC0934)  a background blue galaxy classified Sa with a 
radial velocity equal to 6938 km/s; F12235+0914 is identified as VCC0864 and 
classified as Im or dE.

\section{The FIR to UV flux ratio of individual galaxies}

\subsection{The  FIR to UV flux ratio as an indicator of dust extinction}

\begin{figure}
\resizebox{\hsize}{!}{\includegraphics{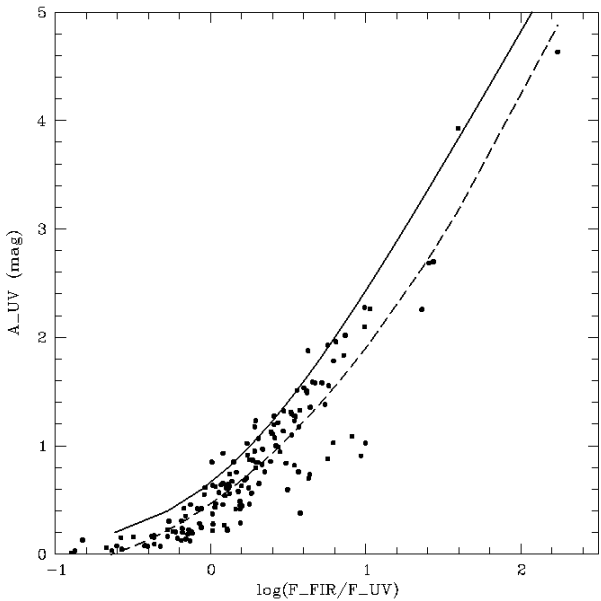}}
\caption[]{The extinction in the UV wavelength range as 
 a function of the ratio of the  FIR and UV  emissions. The points are the 
results of Buat and Xu (\cite{buxu}) and the UV fluxes are taken at 0.2 $\mu$m, 
the dashed curve is the result of the polynomial fit to the points.
The solid curve is obtained using the relation  of Meurer et al. 
(\cite{meurheck}) 
between  the extinction and  the FIR to UV ratio at 0.16 $\mu$m. }
\end{figure}

The FIR to UV ratio in star forming galaxies is now well recognized as a
powerful indicator of extinction.  The basic idea is to perform an energetic
budget:  the amount of stellar emission lost due to the extinction is
re-emitted by the dust in the FIR.  Nevertheless the quantitative calibration
relies on models.  In galaxies with an active star formation activity the
heating of the dust is mostly due to the emission of young and massive stars;
therefore the FIR to UV flux ratio is expected to be tightly related to the
extinction.  Following this approach, Buat and Xu (\cite{buxu}) have estimated
the extinction at 0.2 $\mu$m in star forming galaxies using a radiation 
transfer
model.  Meurer et al.  (\cite{meurheck}) have followed a more empirical way to
relate the FIR to UV ratio to the extinction at 0.16 $\mu$m using a dust screen
model and various extinction curves.  They obtained a relation between the
extinction and the FIR to UV flux ratio and then between the extinction and the
UV slope $\beta$.  \\

The results of both studies are compared in figure 1. The FIR flux is taken in 
the range 40-120 $\mu$m as a combination of the emission at 60 and 100 $\mu$m 
(Helou et al. \cite{helou}) and the UV flux is defined as  $\rm F_\lambda = 
\lambda 
\cdot f_\lambda $ where $\rm f_{\lambda}$ is a flux per unit wavelength.\\
The extinction estimated by the model of Meurer et al. is calculated using 
their 
relation between  the FIR to UV flux ratio and  the extinction: 
$$\rm a_{0.16} = 2.5 \log({F_{FIR}\over{1.19 F_{0.16}}}+1)$$ It is shown as the  
solid curve in figure 1.\\
The results found by Buat and Xu (\cite{buxu}) for their sample of nearby 
galaxies are also reported in figure 1. A polynomial fit on the individual 
points gives: 
$$\rm a_{0.2} = 0.466 (\pm 0.024) + 1.00 (\pm 0.06) \log(F_{FIR}/F_{0.2})$$
$$\rm +0.433 (\pm 0.051) \log(F_{FIR}/F_{0.2})^2$$
The fit is reported as the  dotted curve in figure 1.

  We must account
for the difference in wavelengths used in the two studies. First, we discuss the 
difference between the UV flux of galaxies at 0.2 or 0.16 $\mu$m which affects 
the x-axis of the figure 1.  Deharveng et al.
(\cite{deharveng}) have found the fluxes at 0.165 $\mu$m of a sample of nearby 
galaxies 
systematically higher by 29\% than the fluxes at 0.2 $\mu$m.  Adopting this 
result, the
definition of the UV fluxes as $\rm F_\lambda = \lambda \cdot f_\lambda $ 
almost
cancels the effect of wavelength and we can consider the fluxes at 0.16 and 0.2
$\mu$m as similar.

The extinctions at 0.16 $\mu$m and 0.2 $\mu$m ($\rm a_{0.16}$ and $\rm a_{0.2}$) 
 plotted along the y-axis of the figure 1
can also be considered as similar:  their ratio is expected to vary from 0.9 to
1.1 using the extinction curve of the MW, LMC (Pei \cite{pei}) or that of
Calzetti (\cite{calzetti}).  Therefore we will note both 
values as $\rm a_{UV}$ without any correction.

The UV extinctions derived from the two methods when the FIR to UV flux ratio 
of 
a galaxy is known are tightly correlated
(correlation coefficient 0.99) since both are directly related to this FIR to 
UV flux ratio:  the calculations of Meurer et al.  lead to an
extinction systematically larger than ours, The difference is $\sim  0.2$ mag 
for low FIR to UV flux ratio ($\rm F_{FIR}/F_{UV}<1.5$) and reaches $\sim 0.4$ 
mag for  $\rm F_{FIR}/F_{UV}>5$. This difference may 
arise
from the different assumptions and calculations made in the two studies. Since
they are interested by starburst galaxies Meurer et al.  use a galaxy spectrum
obtained from a constant star formation rate for at most $10^8$ years whereas 
we
use empirical broadband spectra:  the contribution of the old evolved stars to
dust heating is certainly larger in our approach leading to a lower UV
extinction for the same amount of dust emission. 
Another major difference is the treatment of geometrical effects.  Meurer et al.
use a screen model and we calculate the extinction with a radiation transfer
model in an infinite plane parallel geometry where dust and stars are uniformly
distributed and which accounts for scattering effects and disk inclination
(Xu \& Buat  \cite{xubu}).  Finally we assume a Milky Way extinction curve 
whereas Meurer et
al.  adopt a uniform UV extinction for the entire spectrum of the starburst.
Therefore, our model is probably more appropriate for normal star-forming
galaxies and the entire disk of starburst galaxies whereas the calculations of
Meurer et al.  are made for starburst regions.  Given these fundamental
differences and the rather large uncertainties on the corrections for extinction
the two methods are in reasonable agreement.  The FIR to UV flux ratio appears
relatively insensitive to the dust characteristics (type, distribution) and the
stars/dust geometry.  This has been confirmed by the recent study of Witt \&
Gordon (\cite{witt}) who explore various dust distributions (homogeneous or
clumpy), extinction properties (Milky Way or Small Magellanic Cloud) and
stars/dust distributions ( uniform mixture or shells).  Such a robustness makes
the FIR to UV flux ratio a reliable quantitative tracer of the dust attenuation
in star forming galaxies.

\subsection{The variation of the FIR to UV ratio: the influence of the FIR 
selection}

One basic difficulty of these studies based on individual galaxies is that the
samples used are all biased and sometimes in a very complicated sense.  The
diagnostics on the UV slope of nearby galaxies all derive from the compilation
of Kinney et al.  (\cite{kinney}) of IUE observations of starburst galaxies
which is not complete in any sense.  Buat and Xu (\cite{buxu}) have used samples
of star forming galaxies selected on their UV {\it and} FIR emissions leading to
very complicated biases.  Whereas the use of sample of galaxies which may be
strongly biased is probably not a limitation to calibrate the physical link
between the FIR to UV flux ratio and the extinction, the presence of these
biases must be accounted for when generic properties of galaxies are deduced
from these samples.

Our purpose is to use our FIR selected galaxy sample to test the influence of 
such a selection on the
deduced value of the FIR to UV ratio. 
We will consider both fluxes at 60 $\mu$m and in the range 40-120 $\mu$m, the
so-called FIR flux.  Each one  has its own advantages:  on one hand more 
galaxies
have a measured flux at 60 $\mu$m than at 100$\mu$m and the luminosity function
has been derived at 60 $\mu$m, on an other hand the FIR emission over the range
40--120 $\mu$m is more easily related to the total emission of the dust and
hence to the amount of extinction than a single band flux.

In this section, we only discuss the observational biases and therefore use the
data at 60 $\mu$m.  In figure 2 is reported the ratio of fluxes at 60 $\mu$m 
and
0.2 $\mu$m, $\rm F_{60}/F_{0.2}$ as a function of the flux and luminosity of 
the
galaxies at 60 $\mu$m.  $\rm F_{60}$ and $\rm F_{0.2}$ are of the form $\rm
F_\lambda = \lambda \cdot f_\lambda $ where $\rm f_{\lambda}$ is a flux per 
unit
wavelength.  The figure 2a with the flux of the galaxies can be used to
study  the selection bias in limited flux samples.  The figure 2b where are 
reported
the luminosities of the galaxies is useful to discuss the intrinsic properties
of the galaxies.

\begin{figure}
\resizebox{\hsize}{!}{\includegraphics{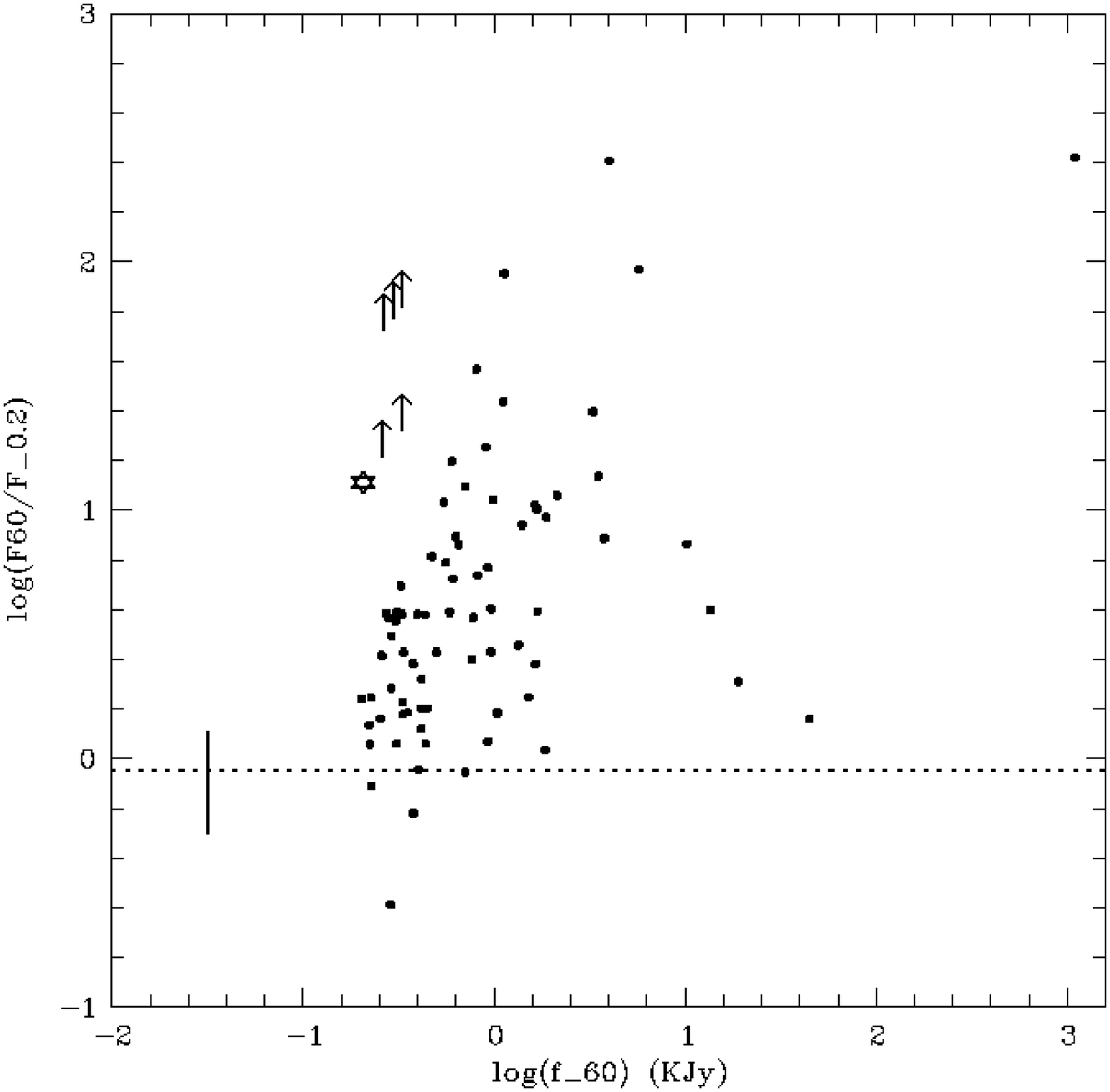}}
\resizebox{\hsize}{!}{\includegraphics{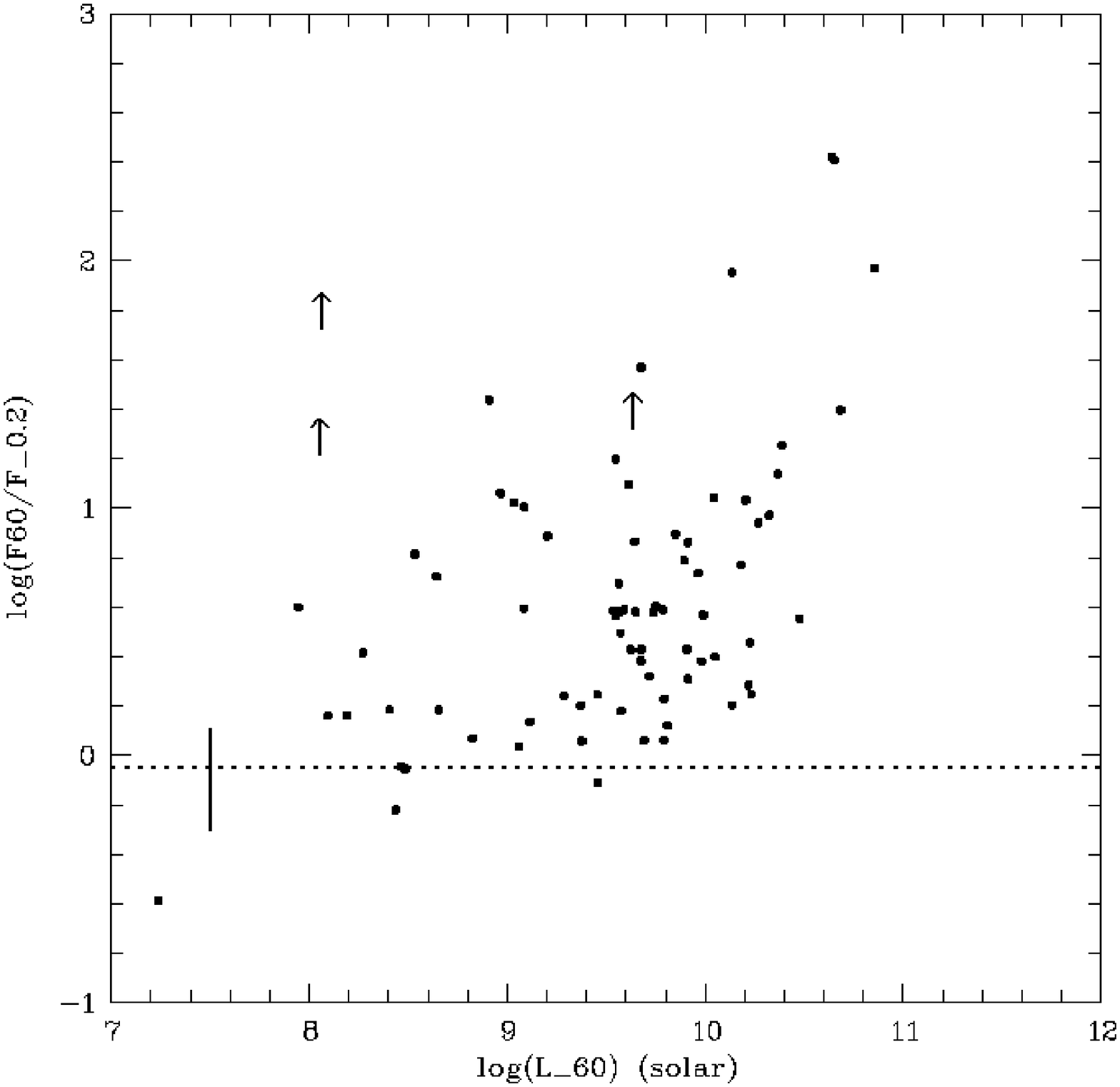}}
\caption[]{The ratio of the emission at 60 and 0.2 $\rm \mu m$ as a function of 
(a)the flux at 60$\rm \mu m$ and (b) the luminosity at 60$\rm \mu m$. The ratio 
of the 
luminosity densities $\rho_{60}/\rho_{0.2}$ is reported as a dotted horizontal 
line, the vertical line is the error bar}
\end{figure}
 
There is a clear trend in both figures in the sense of  a larger $\rm 
F_{60}/F_{0.2}$ ratio
for brighter galaxies at 60 $\mu$m.  The tail found  in figure 2a  at large 60 
$\mu$m flux  toward low $\rm F_{60}/F_{0.2}$ ratios is due to  very nearby 
galaxies. This effect of distance 
disappears 
 when the luminosity is considered (figure 2b). In order to highlight the 
general trend we have
calculated a moving median on the sample.  The data are sorted according to the
60 $\mu$m luminosity, then a  median is calculated for bins of 11 objects
each time shifted by 5 objects.  The result is shown in figure 3.  As expected 
the moving median has reduced the dispersion of the data and flattened the 
dispersed trend of the figure 2b. A linear fit 
gives $$\rm \log(F_{60}/F_{0.2}) = 0.33(\pm 0.09)~ \log L_{60}-2.60(\pm 0.19)$$

\begin{figure} 
\resizebox{\hsize}{!}{\includegraphics{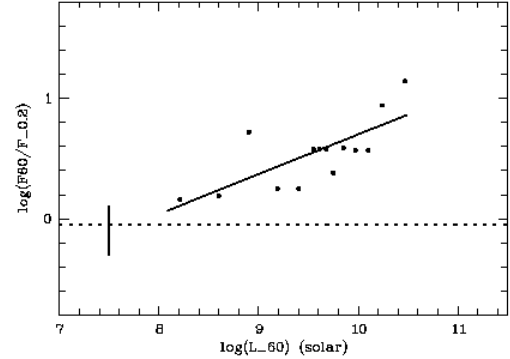}}
\caption[]{The ratio of the emission at 60 and 0.2 $\mu$m as a function of the
luminosity at 60$\mu$ m obtained with a moving median.  The value reported on
the X axis is the mean value of the 60$\mu$m luminosity within each bin.  The 
linear fit is represented by the full line. The
ratio of the luminosity densities $\rho_{60}/\rho_{0.2}$ is reported as a
dotted horizontal line, the vertical line is the error bar} 
\end{figure}

These figures illustrate the bias introduced by a FIR selection. As we consider 
galaxies with an increasing 60 $\mu$m flux or luminosity, their  $\rm 
F_{60}/F_{0.2}$ ratio also increases and is less and less representative of the 
mean properties  of the local Universe as we will see below.

\subsection{The galaxies detected at 60 $\mu$m and not at 0.2 $\mu$m}

The case of these galaxies is especially interesting since they are good 
candidates for very obscured galaxies. Nevertheless their low number (5 cases, 
section 2.2) makes them having no or little influence on the statistical 
properties discussed in this paper. Moreover, little information is known about 
these objects. Only one galaxy (F12242+0919) 
has a known redshift in the NED database.

The  $\rm F_{60}/F_{0.2}$ ratio of each object is reported in table 2 and 
plotted 
in figure 2a. 
The upper limits  found for these galaxies are compatible 
with the values found for some galaxies of the IRAS/FOCA sample but their 
location in the figure is surprising since they do not follow the general 
(although dispersed) trend of a larger $\rm F_{60}$ flux for a larger $\rm 
F_{60}/F_{0.2}$ ratio. However, only the figure 2b where the luminosity of the  
galaxies are reported has a physical meaning and unfortunately only one object 
(F12242+0919) has a measured redshift. Since F12235+0914 and F12259+1141 are 
classified by Yuan et al. as members of the Virgo cluster we assign them a 
distance of 17 Mpc.  These three galaxies are reported in figure 2b. For the 
most 
luminous (F12242+0919) the upper limit of $\rm F_{60}/F_{0.2}$ is compatible   
with  the general trend, the two  faint Virgo dwarfs clearly disagree. Due to 
their faintness not much information is available for them,  F12259+1141 is 
classified as dE and F12235+0914  dE or Im. A large FIR to UV ratio is not 
expected for elliptical galaxies, therefore these objects are probably not dE.
We will see in section 6 that even the most FIR bright and extincted objects 
known in the Universe follow and extend the trend found in figure 2b so the 
 behavior of these two objects is difficult to understand.

We can try to estimate an extinction for the objects listed in table 2.  Only 
two (F12041+6519, F13041+2907)
have been detected at both 60 and 100$\mu$m.   For these two galaxies we have 
the FIR flux to estimate the 
UV 
extinction (a lower limit for  F13041+2907)  using the formula (polynomial fit) 
established in section 3.1.  For
the galaxies not detected at 100 $\mu$m we estimate arbitrarily this flux such
as $\rm f_{60}/f_{100}=0.3$ which is intermediate between the values for warm
and cool dust (Lonsdale \& Helou \cite{lonsdale}), if this value is 
incompatible
with the upper limit, we adopt the upper limit.  The 
extinctions
are listed in table 2.  Adopting the relation of Meurer et al.  leads to 
extinctions larger by 0.4 mag.

Three galaxies have a UV extinction larger than 3.5 mag, they are the two 
objects without any optical identification and the faintest galaxy of the table 
2 detected in B. The three other cases (two non detections and the uncertain 
one) are less extreme ($\rm a_{UV}>\sim 2.5 ~mag$).

Note that the upper limits found for these galaxies are compatible with the
values found for some galaxies of the IRAS/FOCA sample (figures 2).  For 
example
the two most extincted galaxies of our sample, namely M82 and IC732, have a UV
extinction larger than 5 mag and a $\rm F_{60}/F_{0.2}$ ratio larger than 2 in
log unit.

\section{Comparison with the luminosity functions and densities in the 
local universe}

\subsection{The ratio of the luminosity densities}

The luminosity functions and luminosity densities of the local universe are 
available at
both wavelengths (0.2 and 60 $\mu$m).  Therefore we can compare some of their
properties to the characteristics of individual galaxies.  The 60 $\mu$m local
luminosity function and density at z=0 have been calculated by Saunders et al.
(\cite{saunders}).  The 0.2 $\mu$m luminosity function and density have been
derived by Treyer et al.  (\cite{treyer}) at a mean z=0.15.  From these studies
we can calculate the ratio of the local luminosity densities $\rm \rho_{60}
/\rho_{0.2}$ at z=0. To this aim we correct the UV density for the redshift
evolution.  From Madau et al. (\cite{madau}) we estimate that the luminosity
density increases by a factor $\sim 1.5\pm 0.2$ from z=0 to z=0.15 which is 
consistent 
with  the estimates of Lilly et al.  (\cite{lilly}) and Cowie et
al.  (\cite{cowie}).  Applying this factor to the estimate of Treyer et al.  we
obtain $\rm \rho_{0.2}=4.6\pm 2.0 ~10^7~h~L\odot/Mpc^3$.  With $\rm \rho_{60}=
4.2\pm 0.4~10^7~h~L\odot/Mpc^3$ we find $\rm \rho_{60}/\rho_{0.2} = 0.9 \pm 
0.4$
at z=0.  In the same way, from $\rm \rho_{FIR}=5.6\pm 0.6~10^7~h~L\odot/Mpc^3$
(Saunders et al.  \cite{saunders}), we calculate $\rm \rho_{FIR}/\rho_{0.2} =
1.2\pm 0.5$.

$\rm \rho_{60}/\rho_{0.2}$  is
reported in  figures 2 and 3.  The ratio appears lower than almost all the
ratios found for individual galaxies and is systematically lower than all the
median values calculated for increasing 60 $\mu$m luminosity (figure 3).
{\it Therefore the study of individual galaxies of our sample does not lead to 
a
reliable estimate of the mean FIR to UV ratio of the local Universe}.

Our sample is FIR selected since we have searched for FIR galaxies detected in
UV, therefore a bias toward large FIR to UV flux ratio is expected and this
bias increases as we select brighter galaxies (figure 2).  For comparison, we 
can also
re-consider the sample used by Buat \& Xu (\cite{buxu}) :  the galaxies were
primarily selected to have a UV measurement and then searched in the IRAS
database.  Only galaxies detected both in UV and FIR are considered.  Whereas
the selection biases of this sample are very complicated since the primary
selection is on the UV the bias toward the FIR is certainly less strong than
for the IRAS/FOCA sample.

In figure 4 the histograms of $\rm F_{FIR}/F_{UV}$ ratio are reported for 
three
samples:  the IRAS/FOCA sample (solid line), the IUE/IRAS templates of Meurer 
et
al.  (\cite{meurheck}) (dotted line) and the sample of Buat \& Xu (dashed 
line).
We can see that almost all the IRAS/FOCA galaxies and the local IUE/IRAS
templates exhibit a larger ratio
than the ratio of the local luminosity densities $\rm \rho_{FIR}/\rho_{0.2}$.
  The situation is less extreme for the Buat \& Xu sample for which the median 
of the $\rm F_{FIR}/F_{UV}$ flux ratio is 1.66, translating to $\rm 
a_{0.2}=0.71 ~mag$ (0.94 mag with the formula of Meurer et al.). This difference 
in the FIR and UV properties of the samples 
explains the rather low extinction found by Buat \& Xu for this sample as
compared with those obtained by Meurer et al.

The mean property of the local 
Universe in terms FIR to UV luminosity density ratio is not  well represented by 
the samples of galaxies  considered here. 
Therefore much caution must be taken to estimate global correction for 
extinction to be applied to the luminosity function.

\begin{figure}
\resizebox{\hsize}{!}{\includegraphics{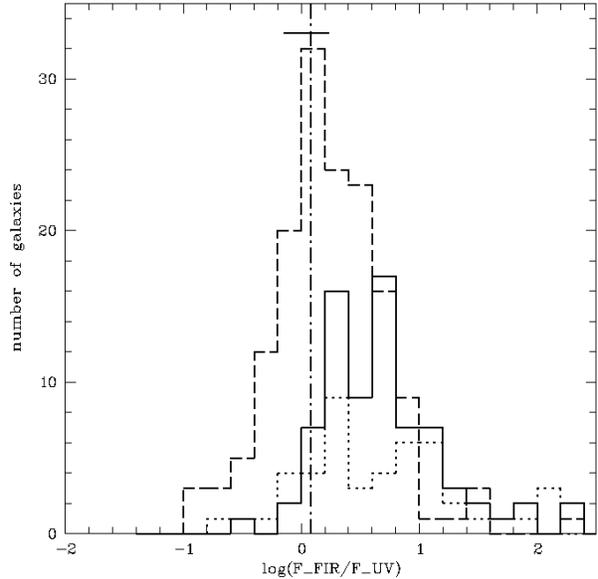}}
\caption[]{The ratio of the FIR and UV fluxes for different samples 
of galaxies:the IRAS/FOCA sample 
(solid line) and the sample of Buat \& Xu (\cite{buxu})  (dashed line) with UV 
fluxes are taken at 0.2 $\mu$m), the IUE/IRAS templates of Meurer et al. 
(\cite{meurheck}) (dotted 
line) with UV fluxes taken at 0.16 $\mu$m. The vertical 
dot-dashed line is the location of $\rm \rho_{FIR}/\rho_{0.2}$, the horizontal 
line is the  error bar}
\end{figure}

\subsection{ The local luminosity functions}

An explanation for the discrepancy between the $\rm F_{60}/F_{0.2}$ ratio of
individual galaxies and the ratio of the local luminosity densities is that it
is not the same galaxies which form the bulk of the UV emission on one hand and
the FIR emission on the other hand.  Indeed, the large difference found in the
shape of the two luminosity functions is consistent with this explanation as
already discussed in Buat \& Burgarella (\cite{bubu}). The adopted value of $\rm 
\rho_{60}/\rho_{0.2}$ depends on the reliability of the luminosity functions 
 and  is subject to some 
uncertainties. Nevertheless very large modifications must be invoked to make 
consistent the $\rm F_{60}/F_{0.2}$ ratios in our IRAS/FOCA sample and the mean 
value of the local universe. Moreover it would not explain the trend found of an 
increase of the $\rm F_{60}/F_{0.2}$ ratio with the FIR luminosity of the 
galaxies. 

We have evaluated the contribution to the luminosity function and the luminosity
density at 0.2 $\mu$m (resp.  60$\mu$m) of the galaxies as a function of their
intrinsic luminosity (per decade of luminosity).  These values are reported in
table 3 (resp 4) together with the number of galaxies of our IRAS/FOCA sample in
each bin of luminosity (in log unit).  The luminosity functions are truncated at
$\rm L=10^7 L\odot$.  

\begin{table*} 
\caption[]{Contribution of the galaxies to
the UV luminosity function and to the UV luminosity density in the local
Universe per decade of luminosity.  The luminosity function is truncated at $\rm
L=10^7 L\odot$($\rm h=0.75$).} 
\begin{flushleft} 
\begin{tabular}{llll} 
$\rm log(L_{UV})$& percentage of galaxies &relative contribution & percentage of
galaxies\\ 
solar unit&from the luminosity function &to the local UV density& in
the IRAS/FOCA sample\\ 
\hline 
7-8& 77.3$\%$&17 $\%$ &11$\%$\\
 8-9& 19.3$\%$&36$\%$&33$\%$\\
 9-10& 3.2 $\%$&36$\%$&56$\%$\\ 
 10-11& 0.2 $\%$&11$\%$&0$\%$\\ 
\hline 
\end{tabular} 
\end{flushleft} 
\end{table*}

\begin{table*}
\caption[]{Same as table 3 at 60 $\mu$m}
\begin{flushleft}
\begin{tabular}{lllll} 
$\rm log(L_{60})$&percentage of galaxies &relative contribution &percentage of 
galaxies\\
solar unit& from the luminosity function &to the local 60$\rm \mu m$ density& in 
the IRAS/FOCA sample\\
\hline
7-8& 45.4 $\%$&2$\%$&3$\%$\\
8-9& 35.6 $\%$&15$\%$&17$\%$\\
9-10&16.2 $\%$&40$\%$&56$\%$\\
10-11& 2.7 $\%$&34$\%$&24$\%$\\
11-12&0.1 $\%$&8$\%$&0$\%$\\
\hline 
\end{tabular} 
\end{flushleft} 
\end{table*}

 As expected for a magnitude limited sample, our individual galaxies do not
truly sample the luminosity functions.  This effect is dramatic in UV:  the
steepness of the faint end slope of the UV luminosity function (Treyer et al.
\cite{treyer}) implies a large number of faint galaxies.  These objects largely
contribute to the UV luminosity density.  The relative numbers of galaxies in
each bin of UV luminosity are similar to those used by Treyer et al.  to 
calculate the
UV luminosity function.\\
Conversely, the FIR luminosity function is better sampled in the sense that the
deficiency of low luminosity galaxies has less implications than in UV.  Indeed,
the FIR luminosity function is extremely flat at low luminosities (Saunders et
al.  \cite{saunders}) and the contribution of the faint FIR galaxies to the
local luminosity density is very low.  As a consequence the number of galaxies
in each bin of luminosity is more representative of its contribution to the FIR
luminosity density than in UV.  The bright end of both luminosity functions is
not represented in the IRAS/FOCA sample because of the scarcity of these objects
and the small statistics.  In terms of global (cumulated) luminosity of our
sample of individual galaxies we are entirely dominated by the galaxies between
$10^9$ and $\rm 10^{11} ~L\odot$ at both wavelengths (0.2 and 60 $\mu$m) but
this does not influence our results since each galaxy is considered individually
whatever its luminosity is, without any summation on individual objects.

Hence our sample IRAS/FOCA sample of individual galaxies is more representative
of the FIR properties of the universe.  If the faint UV galaxies are dwarf
galaxies they probably have a low extinction and therefore a low FIR to UV
ratio.  Our sample being FIR selected, it is probably biased against these
objects.

A consequence of these effects is that when a correction for extinction is
calculated from individual galaxies using such a correction to
correct the entire luminosity function can lead to some mistakes as we will 
discuss in the next subsection.

\subsection {Consequences on the estimate  of the UV extinction for large 
samples of galaxies and statistical  studies}

Most of the time neither the FIR flux nor the UV continuum ($\beta$ slope) are   
available for large
and/or deep surveys of galaxies and one cannot use these dust extinction
calibrators.  The situation is better at high z due to the redshifting of the 
UV
continuum.  For instance  Meurer et al.  ({\cite{meurheck}) have performed 
individual
corrections on the Lyman break U-dropouts galaxies at $z\simeq 3$ in the HDF by
estimating the $\beta$ slope from the V and I measurements.  Nevertheless they
sampled only bright galaxies ($\rm M_{AB}<\sim -19$ i.e.  whose UV luminosity is
larger than $\rm 8 10^9 L\odot$)  since only such bright objects are reachable
at high z.

The problem of the correction for extinction arises when one has to 
derive an intrinsic  UV  luminosity 
distribution (Treyer et al. \cite{treyer}, Steidel et al. \cite{steidel}). 
At low redshift the UV slope is not available for the moment on a large sample
of galaxies and cannot be used to correct the UV luminosity function for dust
extinction. The
extinction has been found  to vary as a function of the absolute bolometric 
magnitude of the
galaxies (e.g.  Wang \cite{wang}, Heckman et al.  \cite{heckrob}, Buat \& 
Burgarella \cite{bubu}).
Unfortunately, the UV luminosity is not a good tracer of the bolometric
luminosity of a galaxy since it is expected to be  very influenced by the 
current star formation
activity. Moreover the extinction (larger for brighter galaxies) adds an anti 
correlation between the observed UV luminosity and the bolometric one.      
Therefore, relating the extinction to the UV luminosity is not
possible. Indeed no correlation exists between the UV luminosity and the
FIR/UV ratio in the IRAS/FOCA sample or that previously used by Buat \& Xu
(\cite{buxu}). In the same way Heckman et al.  (\cite{heckrob}) have used the 
sum of the FIR
and UV luminosities as a tracer of the bolometric luminosity.

The use of the absolute B magnitude $\rm M_B$ is also far from ideal since it 
suffers from the same caveats as the UV luminosity (star formation history and 
extinction), although in a less extreme way. Actually a trend has been found 
between the ratio of the dust to UV emission and $\rm M_B$ for a sample of 
nearby starburst galaxies (Buat \& Burgarella \cite{bubu}) but the relation is 
too dispersed to be used as a quantitative calibrator of the extinction. More 
promising is the use of data at longer wavelengths like the R or I band: the 
effects of extinction will be largely reduced and we can hope to better trace 
the mass of the galaxies. Such investigations  are devoted to a subsequent 
paper. 

We have also compared the extinction deduced from the FIR/UV flux ratio 
($\rm a_{0.2}$, section 
3.1) to the UV-B color since this color is often available for large samples 
(e.g. Treyer et al. \cite{treyer}). The extinction is plotted against the UV-B 
color for our IRAS/FOCA sample in 
figure 5.

A correlation is found between these two quantities (R=0.70).  Indeed, a clear
correlation has already been found between the FIR/UV flux ratio and the UV-B
color (Deharveng et al.  \cite{deharveng}) which has been interpreted to be due
at least in part to the influence of the dust extinction (Buat et al.
\cite{buetal}).  The UV-B color is also sensitive to the star formation history
on timescales of the order of some $10^9$ years:  it is likely to be at the
origin of the dispersion found in figure 5 and only rough tendencies can be
securely deduced.  Nevertheless, it appears necessary to account for the
variation of the extinction among galaxies which can vary by three magnitudes.
In particular galaxies with a UV-B color lower than $\sim -2$ are very little
affected by the extinction and it would seem reasonable to apply no correction
of extinction to them.

\begin{figure}
\resizebox{\hsize}{!}{\includegraphics{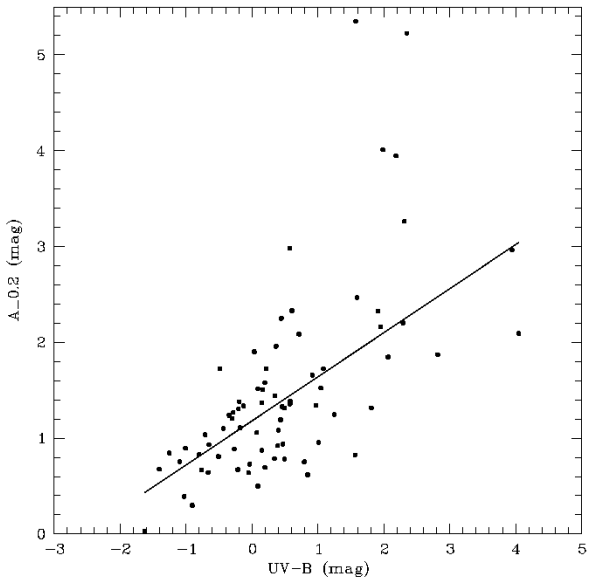}}
\caption[]{The UV extinction at 0.2 $\mu$m calculated with  the FIR to UV flux 
ratio of galaxies as a function of the UV-B color 
 for the IRAS/FOCA sample.
The solid line is the result of a linear regression: $\rm a_{0.2} = 1.18 (\pm 
0.58) + 0.46 (\pm 0.06) (UV-B)$ }
\end{figure}

\section{Estimating star formation rates}

Both FIR and UV emissions are powerful star formation tracers.  To derive
reliable star formation rates (SFRs) one must account for the repartition of 
the
emission of young stars in both wavelength ranges since the stellar emission
lost by dust extinction is re-emitted in the FIR.

As already proposed by Heckman et al.  (\cite{heckrob}), perhaps the best way 
is
 to consider both UV and FIR emissions:  each emission can be related to the
 star formation rate and the sum of the two SFRs deduced from the FIR and the 
UV
 should account for the total emission of young stars.\\ 
In such an approach the uncertainty resides in the translation of the UV and FIR
emissions into quantitative star formation rate.  The UV flux is directly
proportional to the star formation rate provided that the star formation has
been constant for some $10^8$ years and assuming a universal initial mass
function (IMF).  We have used the models of Leitherer et al.  (\cite{leitherer})
for different IMFs (Salpeter IMF (-2.35) or -2.5, upper mass limit 100 or 125
M$\odot$ for a lower mass limit equal to 0.1 M$\odot$).  The metallicity is
taken solar.  After 5 $10^8$ years of constant star formation the production of
the UV luminosity reaches stationarity :  $\rm SFR (M\odot/yr) = 2.7~10^{-10}
L_{UV} (L\odot)$ at 0.2 $\mu$m for a Salpeter IMF and an upper mass limit of 125
M$\odot$.  Nevertheless after 5 $10^7$ years the UV luminosity reaches more than
80$\%$ of this stationary value.  For the same IMF, using Madau estimations
(\cite{madau}) based on the models of Bruzual \& Charlot Treyer et al.
(\cite{treyer}) have adopted a conversion factor $\rm SFR/L_{UV}$ equal to
$3.36~10^{-10} M\odot yr^{-1}/L\odot$ i.e.  a difference of $20 \%$.  The
uncertainty due to the IMF is around $\sim 40\%$.  Therefore we can
conservatively estimate that the uncertainty on the conversion of the UV
luminosity in star formation rate is $50 \%$ provided that the galaxy has formed
stars continuously for some $10^7$ years.\\
The link between the SFR and the FIR luminosity is more indirect than for the UV
luminosity since it depends of the dust heating which involves all types of
stars.  Nevertheless, in starbusting galaxies the situation is expected to be
less complex since the dust heating is dominated by the young stars.  Under such
conditions Kennicutt (\cite{kennicutt}) has related the FIR luminosity to the
star formation rate $ \rm SFR=1.71~10^{-10}~ L_{FIR} (L\odot)$ where $ \rm
L_{FIR}$ is the total FIR luminosity.  This conversion factor $\rm SFR/L_{FIR}$
is obtained using synthesis population models and is also subject to
uncertainties on the stellar tracks, the IMF or the star formation history.
From a comparison with the calculations of Lehnert \& Heckman ({\cite{lehnert}), 
Meurer et al.  (\cite{meurheckI}) and Buat \& Xu (\cite{buxu}) we estimate the 
uncertainty of the order
of 50$\%$ for $\rm SFR/L_{FIR}$.\\

Therefore we can reasonably estimate that the SFR deduced from the observed UV 
luminosity added to that deduced from the FIR one is also uncertain by a factor 
$\sim 50\%$. Nevertheless it must be noticed that the conversion formulae only 
apply to galaxies which have experimented a continuous star formation for at 
least $\sim 10^8$ years and will not be valid for galaxies with more episodic 
star formation, especially post starbursting galaxies.

\subsection{The IRAS/FOCA sample}

The IRAS/FOCA sample is FIR selected, thus it is biased against very blue dwarf 
galaxies which may exhibit episodic bursts of star formation as suggested by 
Fioc \& Rocca-Volmerange (\cite{fioc}). Therefore we can expect that the 
derivation of a SFR from the FIR and UV emissions is valid for this sample of 
galaxies. 
 The comparison of the
star formation rates obtained by adding the  FIR and UV (observed) emissions to  
 those deduced from the UV 
fluxes
after  a correction for extinction can be useful to test the consistency of
both methods. 
Therefore we have calculated the star formation rates for the IRAS/FOCA sample 
 of galaxies adding the contribution of the FIR and UV (not corrected for 
extinction) emissions 
and using the  conversion formula of Treyer et al. (\cite{treyer}) for the UV 
and Kennicutt (\cite{kennicutt}) for the FIR. Total FIR fluxes have been 
estimated using the relation found by Buat \& Burgarella (\cite{bubu}) between 
the ratio of the total dust flux to the FIR (40-120$\mu$m) flux and  $\rm 
f_{60}/f_{100}$. The  estimated SFRs can be  
 compared  to the ones obtained after  correction of  the UV 
fluxes from extinction. The correlation between the two estimates  is 
very good but the SFRs deduced from the ($\rm FIR+UV$) emissions are higher 
than 
the SFRs deduced from the UV corrected emission by a factor 1.4. Another 
interesting comparison is that of the relative contribution of the UV (not 
corrected for extinction) and FIR 
emissions to the total ($\rm FIR+UV$) SFR. For our sample of IRAS/FOCA 
galaxies the relative contribution of the UV and FIR emissions to the total  
SFR are 0.3 and 0.7 respectively. However, these calculations assume that  the 
FIR flux is exclusively due to the heating by young stars. Since all our 
galaxies are certainly  not  starbursting objects the contribution of the 
emission of dust heated by old evolved stars must be deduced from the FIR flux 
before translating it into star formation rate reducing the contribution of the 
FIR emission to the SFR. Let us assume that old stars contribute for 30 $\%$ of 
the dust heating (Xu \cite{xu}, Buat \& Xu \cite{buxu}), then the ratio of the 
SFR deduced from the ($\rm 
FIR+UV$) emissions and the SFR deduced from the UV corrected emission is  
reduced from 1.4 to 1.1 and the relative contributions of the UV and FIR 
emissions to the total  SFR are now  0.4 and 0.6.

 Given all the uncertainties inherent to these calculations we must be cautious 
in our conclusions. We can say that the corrections for dust extinction deduced 
from the FIR/UV flux ratio and applied to the UV observed emissions  lead to a 
SFR 
 consistent with that obtained by adding the SFRs deduced from  both the FIR 
and 
observed UV emissions. This makes us confident in our estimate of the 
extinction.

 \subsection{The local volume-average star formation rate }
 
 Under the hypothesis of an average  star formation of the local universe  
continuous over $\sim~10^8$ years, we 
  can  derive global star formation rates from the local FIR and UV 
luminosity densities: 
 
 $$\rm \rho_{SFR} = \rho_{SFR}^{FIR} +\rho_{SFR}^{UV}$$
 
 $\rm \rho_{SFR}^{FIR}$ is calculated using the value of $\rm \rho_{FIR}$ 
(section 4.1) multiplied by 1.5 to account  for the total dust luminosity 
density (e.g. Xu \& Buat \cite{xubu}, Meurer et al. \cite{meurheck}); $\rm 
\rho_{SFR}^{UV}$ is calculated from $\rm\rho_{0.2}$. We find: 
 
 $$\rm  \rho_{SFR} =  (0.014  +0.015) (\pm 0.01)   ~ ~h \cdot M\odot/yr/Mpc^3$$

 This time the contributions of the FIR and UV are very similar, this is due to 
the lower value of the FIR to UV density ratio as compared to the FIR to UV 
flux 
ratio of individual galaxies.  Then the volume-average star formation rate 
deduced from the UV luminosity density not corrected for dust extinction must 
be 
multiplied by a factor $\sim  2$ to account for the global extinction, this 
corresponds 
to a mean extinction of 0.75 mag at 0.2 $\mu$m. As discussed before this rather 
low value is due to the contribution of faint blue galaxies to the UV 
luminosity 
density.\\
Actually, using the $\rm \rho_{FIR}/\rho_{0.2}$ ratio gives an extinction of 
0.77 mag using the model of Meurer et al. and 0.55 mag for our polynomial fit 
(section 3.1)
As already underlined, the difference is likely to come from the contribution 
of 
the old stars to the dust heating: let us assume that 30$\%$ of the FIR 
emission 
comes from  these old stars and is not related to the recent star formation 
then 
 we find that $\sim 40\%$ of the star formation is locked in FIR and $60\%$ in 
UV. The resulting UV extinction is 0.58 mag and $\rm \rho_{SFR}~ =~ 
0.025~h~M\odot/yr/Mpc^3$.

Therefore we can conclude that the derivation of the  global star formation 
rate 
is in agreement with our estimate of the global  extinction in UV and that the 
same amount of star formation rate is traced by the global FIR and UV (not 
corrected for extinction) luminosity densities. 

\section{Comparison with  FIR bright galaxies }

Since the IRAS survey, FIR bright galaxies have been the subject of numerous 
studies because these objects experiment an intense star formation activity. 
The 
extreme case is that of UltraLuminous Infrared Galaxies (ULIGs) with a 
bolometric luminosity larger than $\rm 10^{12} M\odot/yr$ essentially emitted 
in 
FIR and star formation rates of several hundreds solar masses per year: they 
generally are violent mergers and may represent an important phase in the 
formation of large galaxies like ellipticals (Mirabel \& Sanders 
\cite{sandmir}).
Such objects are known to be rare at low z but they might be far more numerous 
at high z as suggested by the sub millimetric surveys with SCUBA (e.g. Sanders 
\cite{sanders}). 

With the launch of the ISO satellite, the sensitivity of the ISOCAM camera has 
allowed  mid-infrared surveys at intermediate redshift ($\rm z<1$). In 
particular 
Flores et al. (\cite{flores}) have observed one CFRS field, therefore UV 
(0.28$\mu$m) and infrared data are available for these galaxies. 

We now compare  the FIR and UV properties of these galaxies (ULIG and 
ISOCAM/CFRS) to that of our IRAS/FOCA sample of nearby galaxies. The comparison 
is rather straightforward since all these objects are IR selected. 

\subsection{Ultraluminous Infrared Galaxies (ULIG)}
 
\subsubsection {nearby ULIGs}

Trentham et al.  (\cite{trentham} ) have obtained HST observations for three
ultra luminous infrared galaxies:  VII Zw031, IRAS F12112+0305, IRAS
F22491-1808.  These galaxies are selected to be cool in order to avoid a non
thermal origin for the FIR emission.  We can calculate directly their $\rm
L_{60}/L_{UV}$ ratio using the data at 0.23 $\mu$m for the UV emission.  The
three objects are reported in figure 6 (similar to fig.2b) with empty stars for
symbols.  As expected for this type of objects they appear to be very luminous
at 60 $\mu$m with a high FIR to UV flux ratio.  Such objects are not represented
in our FIR selected sample of nearby galaxies:  this emphasizes how much these
objects are rare in the local Universe and with extreme properties as often
underlined (e.g.  Sanders \& Mirabel \cite{sanders}).  Since the three ULIG have
also been detected at 100 $\mu$m we can estimate their UV extinction (we neglect
the difference in the UV wavelengths i.e.  0.23 $\mu$m versus 0.2 $\mu$m).  We
find $\sim 6.5$ mag:  more than 99$\%$ of the UV flux of these objects is
emitted in the FIR.

\subsubsection{ High redshift galaxies detected by SCUBA : ULIG candidates}

Hughes et al.  (\cite{hughes}) have observed the HDF field at 850 $\mu$m with
SCUBA.  5 objects detected by SCUBA in the HDF field have been tentatively
associated to optical sources for which photometric redshift are available but
such an identification is difficult because of the uncertainty on the 850 $\mu$m
positions.  Indeed, the identification of the most brightest source (HDF850.1)
has not been confirmed (Sanders, \cite{sanders}).  Moreover the nature of these
sources, starbursts or AGN, is not clear:  at FIR luminosity larger than $\rm
10^{12}~L_{\odot}$ about half of the nearby ULIGs are predominantly powered by
AGNs (e.g.  Sanders \cite{sanders}).  \\

The 60 $\mu$m luminosity of the 4 remaining galaxies is obtained from their 
emission at
850 $\mu$m accounting for the redshifting and an assumed spectral energy
distribution chosen to be that of M82.  The optical data from the HDF lead to
the estimate of the UV flux at a rest frame wavelength of 0.28 $\mu$m.  All 
these
estimates rely on the resemblance of all ULIGs with M82 and can lead to false  
results (e.g. Sanders \cite{sanders}) . In spite of these caveats, 
 we have reported the 4 high redshift galaxies in figure 6 (filled triangles). 
They appear very extreme, being more luminous in FIR and probably more extincted 
that 
all the other galaxies studied in this paper. A tentative estimate of the 
extinction is obtained by 
using the $\rm F_{60}/F_{UV}$ instead of the $\rm F_{FIR}/F_{UV}$ one. We find 
values spanning from 8 to 11 mag. As a comparison M82, which belongs to our 
IRAS/FOCA sample exhibit "only" 5.4 mag of extinction at 0.2 $\mu$m.
These high redshift ULIGs seem also to be  much more extincted than the most 
luminous Lyman break galaxies of the HDF studied by Meurer et al. 
(\cite{meurheck}) for which they derive an extinction not larger than 3.5 mag.
 Although their UV  luminosity corrected for extinction  are comparable ($\rm 
\sim 
10^{12}~ 
L\odot$), these two classes of galaxies do not seem to exhibit the same 
properties in FIR and UV as suggested by Heckman (\cite{heckman}). Indeed we 
can 
try to roughly locate the most luminous galaxies of Meurer et al. in figure 6. 
The FIR luminosity can be estimated from their star formation rates and the 
$\rm F_{60}/F_{UV}$ ratio from their extinction using the figure 1. It gives 
$\rm L_{60}\sim 10^{12} L\odot$ and $\rm F_{60}/F_{UV}\sim 1.2-1.4$ for an 
extinction of $\sim 3$ mag. Therefore it seems that the  Lyman Break Galaxies 
detected in the HDF by their U-dropout do not follow the steep trend of figure 
6 
found for FIR bright galaxies but instead exhibit a lower increase of the 
extinction with the intrinsic luminosity of the galaxies. Such a difference may 
be due to the contribution of  AGNs in ULIGs. Indeed the extrapolation of the 
mean trend found in 
the IRAS/FOCA sample (figure 3) reported as a full line in the figure 6 does not 
lead to the extreme case of ULIGs and seems more compatible with LBGs.

\subsection{ ISOCAM/CFRS galaxies}

Flores et al.  (\cite{flores} ) have obtained ISO/ISOCAM Mid Infrared  images 
of 
one CFRS
field, most of the detections are at 15 $\mu$m.  The infrared 8-1000 $\mu$m
luminosities have  been deduced from MIR and/or radio measurements using
templates of spectral energy distributions and are probably not very secure but
an approximate value is sufficient for our comparison with local templates. We 
 differentiate AGNs and starbursts  as classified by Flores et al. 
Only the global IR (8-1000 $\mu$m) flux is available for these objects and not 
the flux at 60
$\mu$m. We adopt a mean value of $\sim 1.5$ for the ratio of the total
dust emission to that intercepted by the 40-120 $\mu$m band (see section 5.2).
 Then we  estimate the ratio between the flux at 60 $\mu$m and the FIR
(40-120 $\mu$m) for our IRAS/FOCA sample:  $\rm L_{60}/L_{FIR} = 0.84\pm
0.12$.  
Therefore the total IR fluxes given by Flores et al.  have been divided by
a factor 2 in order to roughly represent the flux at 60 $\mu$m.\\
The UV emission is taken at 0.28 $\mu$m as given by  Flores et al. It is 
difficult to estimate a correction factor to translate the UV data to 0.2$\mu$m 
in the absence of observations of a large sample of galaxies at both 
wavelengths since the ratio depends on the star formation history and the dust 
extinction. We can try to use synthesis models for this estimate:  assuming a 
constant star formation rate over 1 Gyr  and using the models of Leitherer et 
al. (\cite{leitherer}) for a solar metallicity we find $\rm 
F_{0.2}/F_{0.28}=1.7$ the flux being defined as $\rm \lambda\cdot 
f_{\lambda}$.The 
difference of extinction between 0.28 and 0.2 $\mu$m has been calculated using 
the extinction curves of the Milky Way and the LMC (Pei \cite{pei}) and that of 
Calzetti (\cite{calzetti}). The ratio $\rm A_{0.2}/A_{0.28}=1.2-1.4$. Therefore 
the two effects (star formation and extinction) roughly compensate each other 
and  we do not perform any correction between 0.28 and 0.2 $\mu$m.
The galaxies are plotted in figure 6 as crosses for the true starbursts 
and empty circles for the Seyferts. They all fill the gap between  the IRAS/FOCA 
sample and  the ULIGs. Therefore, 
they are not as extreme as ULIGs but their extinction is larger than the nearby 
galaxies of the IRAS/FOCA 
sample. For the galaxies classified as starbursts, we have tentatively estimated 
this extinction  from
 their FIR to UV flux ratio.  The 
extinctions found span from 2 to 5.5 with a 
mean at 3.3 mag 
(and a median at 3 mag). This is much larger than that estimated by Flores et 
al. by matching the global star formation rates deduced from the total FIR and 
the UV 
luminosities in the observed field: they find extinctions  around 2 mag at 
0.28 
$\mu$m. This discrepancy between the extinction occurring in individual 
galaxies 
selected in infrared and that deduced from the total FIR and UV luminosity of a 
selected field (i.e. the sum of the luminosity of all galaxies detected in the   
wavelength band (FIR or UV)) is well illustrated in the table 5 of Flores et 
al. 
where the ratio $\rm L_{IR}/L(0.28\mu m)$ calculated for individual objects 
observed at 
both 15 and 0.28 $\mu$m   is  $\sim 5$ times larger than  the ratio of the 
global luminosities IR and UV luminosities in the CFRS field. This is in full 
agreement with our own results presented in section 4.

\begin{figure}
\resizebox{\hsize}{!}{\includegraphics{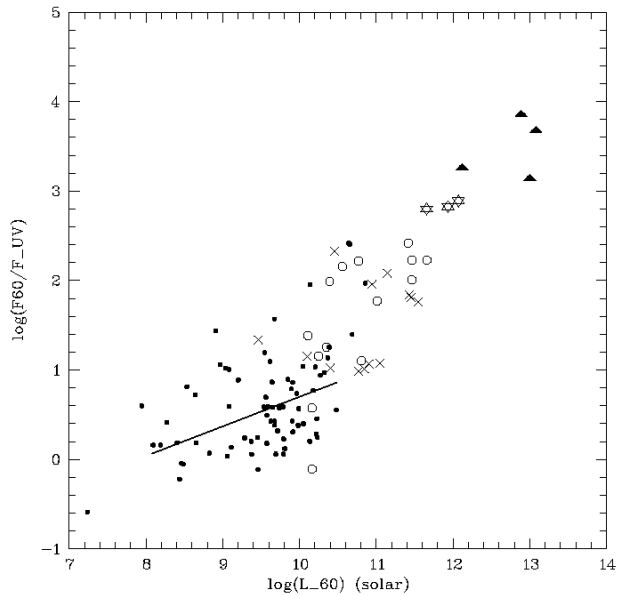}}
\caption[]{FIR bright galaxies are superposed to our IRAS/FOCA sample (dots): 
 ISOCAM/CFRS starburst galaxies with crosses, ISOCAM/CFRS active galaxies 
with empty circles, nearby ULIGs with stars and distant 
SCUBA detections with filled triangles (h=0.75).}
\end{figure}

\section{Conclusions}

We have constructed a sample of 102  nearby galaxies detected  by IRAS at 60 
$\mu$m and for which UV observations at 0.2 $\mu$m are available down to $\rm 
m_{UV}\sim 17-18$ . Only five galaxies have no UV detection implying an 
extinction larger than 2-3 mag for these objects which are also very faint in 
FIR. 

The FIR and UV properties of our sample have been compared to the mean 
properties of
the local Universe deduced from the luminosity functions and densities at both
wavelengths.  As the galaxies become brighter in FIR their FIR to UV flux 
ratio,
i.e.  their extinction increases:  $\rm d(\log( L_{60}/L_{0.2}))/d(\log 
L_{60})\simeq 0.3$  which translates to an increase of $\rm \sim 0.5$ mag for 
the 
dust extinction in UV per decade of FIR luminosity.

The ratio of the FIR to UV local luminosity
densities is much lower than that found in individual galaxies.  It is also 
true
for other samples of nearby galaxies usually considered as low redshift
templates like the IUE sample of Calzetti Kinney and collaborators.  Such a 
difference
is likely to be due to the large contribution of low UV luminosity galaxies to
the UV luminosity density:  these galaxies are deficient in  any survey.  At FIR 
wavelengths
such faint galaxies do not significantly contribute and our sample is 
more representative of the  galaxy population in terms of its contribution to 
the FIR luminosity density.\\
As a consequence, much caution must be taken to correct large samples of 
galaxies for  extinction. In particular a uniform correction deduced from the 
study of some individual cases cannot be valid.  

Star formation rates can be estimated by accounting for both the FIR and UV
emissions:  each one is translated to a quantitative SFR; then, the two SFRs 
are summed. 
The SFRs such deduced  are consistent with those calculated from the UV  
emission corrected for extinction.\\
A local volume-average star formation rate is calculated from the FIR and UV 
luminosity density: $\rm \rho_{SFR} = 0.03 \pm 0.01 ~h \cdot
M\odot/yr/Mpc^3$. This is consistent with a global extinction of $\sim 0.6$ mag 
at 0.2 $\mu$m.

Endly, we have compared the FIR and UV properties of our sample of galaxies to
those of nearby and high redshift UltraLuminous Infrared Galaxies observed both
at UV and FIR rest frame wavelengths and to the ISOCAM survey of a CFRS field.
All these objects extend toward the large luminosities the trend found for the
nearby galaxies of a larger FIR to UV ratio for brighter galaxies.  The ULIGs
are very extreme with UV extinctions reaching 8-11 mag.  Although more moderate
the extinctions we find for the ISOCAM/CFRS objects not classified as Seyfert 
are comprised between 2 and
5.5 mag.  These calculations are only tentative due to the large uncertainties
about the  FIR emission of these objects.

\begin{acknowledgements} 
We thank J.-M. Deharveng and M. Treyer for their careful reading of the 
manuscript and  A.  Boselli 
for 
helpful and stimulating
discussions about this work as well as for his help in the study of individual
objects.

\end{acknowledgements}

\end{document}